\documentclass[%
 reprint,
 superscriptaddress,
 amsmath,amssymb,
aps,
pra,
]{revtex4-1}

\usepackage[colorlinks, citecolor=red]{hyperref}
\usepackage{graphicx}
\usepackage{dcolumn}
\usepackage{bm}

\usepackage{listings}
\usepackage{color}
\usepackage{subfigure}

\definecolor{dkgreen}{rgb}{0,0.6,0}
\definecolor{gray}{rgb}{0.5,0.5,0.5}
\definecolor{mauve}{rgb}{0.58,0,0.82}

\begin{document}


\title{Effects of Quantum Noise on Quantum Approximate Optimization Algorithm}
\author{Cheng Xue}
 \affiliation{Key Laboratory of Quantum Information, Chinese Academy of Sciences, School of Physics, University of Science and Technology of China, Hefei, Anhui, 230026, P. R. China}
\author{Zhao-Yun Chen}
 \affiliation{Key Laboratory of Quantum Information, Chinese Academy of Sciences, School of Physics, University of Science and Technology of China, Hefei, Anhui, 230026, P. R. China}
 \affiliation{Origin Quantum Computing Hefei, Anhui 230026, P. R. China}
 
\author{Yu-Chun Wu}
 \affiliation{Key Laboratory of Quantum Information, Chinese Academy of Sciences, School of Physics, University of Science and Technology of China, Hefei, Anhui, 230026, P. R. China}
\author{Guo-Ping Guo}
 \email{gpguo@ustc.edu.cn} 
 \affiliation{Key Laboratory of Quantum Information, Chinese Academy of Sciences, School of Physics, University of Science and Technology of China, Hefei, Anhui, 230026, P. R. China}

\begin{abstract}

    The quantum-classical hybrid algorithm is an algorithm that holds promise in demonstrating the quantum advantage in NISQ devices. When running such algorithms, effects from quantum noise are inevitable. In our work, we consider a well-known hybrid algorithm, the quantum approximate optimization algorithm (QAOA). We study the effects on QAOA from typical quantum noise channels and produce several numerical results. Our research indicates that the output state fidelity, the cost function, and its gradient obtained from QAOA decrease exponentially with respect to the number of gates and noise strength. Moreover, we find that noise merely flattens the parameter space without changing its structure, so optimized parameters will not deviate from their ideal values. Our result provides evidence for the effectiveness of hybrid algorithms running on NISQ devices.

\end{abstract}

\maketitle


\section{Introduction}

In recent years, there have been rapid developments towards building quantum computers. The expectation is that they will have a quantum chip containing hundreds of qubits---in essence, a noisy-intermediate-scale quantum (NISQ)\cite{NISQ_preskill_quantum_2018} device. However, some quantum algorithms will require enormous numbers of qubits and a quantum circuit having a long circuit depth to realize quantum supremacy\cite{preskill_quantum_2012}. When running complex quantum circuits, quantum error correction codes need to be executed, but this will further increase the required number of qubits and quantum circuit complexity. For example, to factor 2048-bit RSA integers requires 20 billion noisy qubits\cite{gidney_how_2019}. This requirement far exceeds the capacity of current quantum chips and will remain so for the near future.

One meaningful problem is to find a quantum algorithm that can realize quantum supremacy on NISQ devices. A quantum-classical hybrid algorithm has promise. In recent years, several quantum-classical hybrid algorithms have been proposed\cite{qcl_2018,havlicek_supervised_2019,QGAN_lloyd, QGAN_PRA_2018, q_autoencodrs, q_BoltzmannMachine, farhi_classification_2018, schuld_quantum_2018, schuld_circuit-centric_2018, schuld_circuit-centric_2018}, a few of which have been demonstrated experimentally on real quantum chips\cite{schuld_implementing_2017,havlicek_supervised_2019,hamilton_generative_2018,zhu_training_2018,grant_hierarchical_2018,tacchino_artificial_2019,rocchetto_experimental_2019,ding_experimental_nodate,benedetti_generative_2019}. In NISQ devices, quantum noise is inevitable, and quantum algorithms running on NISQ devices need to be noise tolerant. However, 
little work has been done to study how quantum noise affects the performance of these quantum-classical hybrid algorithms. In \cite{sharma_noise_2019}, Sharma at al studied the effects of quantum noise on variational quantum compiling\cite{khatri_quantum-assisted_2019,jones_quantum_2018,heya_variational_2018} and proposed optimal parameter resilience (OPR).

We focus on a specific quantum-classical hybrid algorithm---the QAOA \cite{qaoa_2014,qaoa_rigetti_2017,qaoa3_2018,qaoa2_2018,qaoa1_2018}. There is little work concerning the effects of quantum noise on the QAOA. In our work, we study specifically the effect of quantum noise on its performance. We analyze the effect of quantum noise on various aspects of QAOA, including the fidelity of its output state, its cost function, its cost function gradient, and its parameter optimization. We find that QAOA is noise tolerant although the QAOA parameter space is somewhat flattened by noise. Nevertheless, the architecture of the parameter space remains basically unchanged, and therefore quantum noise does not affect the optimization process of the QAOA parameters. To demonstrate our conclusion, we choose three different quantum noise channels and test the effect of each on QAOA through numerical simulations.

\section{Preliminary}

\subsection{QAOA}\label{QAOA_method}

The QAOA was developed by Farhi, Goldstone, and Gutmann\cite{qaoa_2014}. It gives a way to solve approximately in polynomial time combinatorial optimization problems such as the Max-Cut problem, the traveling salesman problem, and the 3-SAT problem .

The QAOA can be regarded as a simplified version of an adiabatic evolution, which involves the continuous evolution of a quantum state from a ground state of a Hamiltonian to a ground state of another Hamiltonian. As the QAOA is discrete, it drives the ground state of one Hamiltonian to the ground state of another Hamiltonian in discrete steps. In the respect, the QAOA is an approximation algorithm. In general, the output state of the QAOA is just close to the ground state of the target Hamiltonian. The QAOA quantum circuit has a parameterization. Changing the parameters of the circuit adjusts the direction of each step of the QAOA. In particular, the QAOA quantum circuit can be written
\begin{equation}
U(\vec{\gamma},\vec{\beta})=\prod_{j=1}^{n}{e^{-iH_d\beta_j}e^{-iH_p\gamma_j}},
\end{equation}
where $n$ denotes the QAOA step number, and $H_p$ the problem Hamiltonian. For a specific problem, the ground state of $H_p$ is the solution of the problem. $H_d$ is the drive Hamiltonian of the QAOA, the initial state of the QAOA being the ground state of $H_d$. Generally, we define $H_d$ in the form
\begin{equation}
H_d=\sum_i{-X_i},
\end{equation}
where $X$ represents the Pauli X matrix. The initial state of the QAOA is then $|\phi_{in}\rangle=|+\rangle^{\otimes m}$, $m$ being the qubit number of the QAOA quantum circuit, and $\vec{\gamma}$ and $\vec{\beta}$ the two $[n,1]$ parameter vectors. The purpose of the QAOA is to find the optimal values of $\vec{\gamma}$ and $\vec{\beta}$ that ensures the output quantum state is closest to the ground state of the problem Hamiltonian $H_p$. Generally, we use a classical optimization algorithm to find the optimal $\vec{\gamma}$ and $\vec{\beta}$. The cost function is the expectation of $H_p$ and may be written
\begin{equation}\label{cost_function}
f(\vec{\gamma},\vec{\beta})=\langle\phi_{in}|U^{\dagger}(\vec{\gamma},\vec{\beta})
H_pU(\vec{\gamma},\vec{\beta})|\phi_{in}\rangle.
\end{equation}

When solving specific problems, the whole execution of the QAOA comprises five steps:
\begin{itemize}
\item [(1)] Construct problem Hamiltonian $H_p$. The ground state of $H_p$ is the solution of the problem.
\item [(2)] Use $H_p,H_d$ to construct the QAOA quantum circuit---the initial values of $\vec{\gamma}$ and $\vec{\beta}$ may be chosen randomly.
\item [(3)] Execute QAOA quantum circuit multiple times and compute the cost function $f(\vec{\gamma},\vec{\beta})=\langle H_p \rangle$.
\item [(4)] Optimize $\vec{\gamma}$ and $\vec{\beta}$ to minimize the cost function $f(\vec{\gamma},\vec{\beta})$ with the classical optimization algorithm.
\item [(5)] Execute the optimized QAOA quantum circuits multiple times---the optimal result is the solution of the problem.
\end{itemize}

\subsection{Quantum noise model}\label{NOISE_MODEAL}

In our numerical simulations, we assess the effects of three quantum noise channels on QAOA, namely, dephasing, bit flip, and depolarizing channels. The Kraus operators of each are given below:\\
Dephasing noise:
$$
K_1=\left[ \begin{array}{ccc}\sqrt{1-p} & 0 \\
0 & \sqrt{1-p}\end{array} \right ],
K_2=\left[ \begin{array}{ccc}\sqrt{p} & 0 \\
0 & -\sqrt{p} \end{array}\right ]
$$
Bit flip noise:
$$
K_1=\left[ \begin{array}{ccc}\sqrt{1-p} & 0 \\
0 & \sqrt{1-p}\end{array} \right ],
K_2=\left[ \begin{array}{ccc} 0& \sqrt{p} \\
\sqrt{p} & 0 \end{array}\right ]
$$
Depolarizing noise:
$$
K_1=\sqrt{1-\frac{3}{4}p}\left[ \begin{array}{ccc}1 & 0 \\
0 & 1\end{array} \right ],
K_2=\frac{\sqrt{p}}{2}\left[ \begin{array}{ccc} 0 & 1 \\
1 & 0 \end{array}\right ]
$$
$$
K_3=\frac{\sqrt{p}}{2}\left[ \begin{array}{ccc} 0 & -i \\
i & 0 \end{array}\right ],
K_4=\frac{\sqrt{p}}{2}\left[ \begin{array}{ccc}1 & 0 \\
0 & -1 \end{array}\right ]
$$

These three channels have one parameter $p$. We choose $p$ in the range $[0.0001,0.02]$; more specifically, we focus on a smaller interval and therefore used the exponential interval,
\begin{equation}
p_i=0.0001*200^{0.1\times i}, i=0,1,...,10.
\end{equation}

\section{Effects of quantum noise on QAOA}\label{EFFECT_ON_QAOA}
Now we analyze the effects of quantum noise on QAOA output state, cost function, cost function gradient, and parameter optimization process in turn. In our analysis, we consider quantum noise for which the Kraus operators are written 
\begin{equation}
K=\{a_0I,a_1K_1,a_2K_2,...,a_sK_s\},
\end{equation}
each $a_i$ being a real number and $I$ the identity operator. We test the effect of this quantum noise on the QAOA performance in a treatment of the maximum cut (Max-Cut) problem in graph theory.

\subsection{State fidelity}\label{Inf_on_state}

As shown in Sec.~\ref{QAOA_method}, for a specific QAOA quantum circuit $U(\vec{\gamma},\vec{\beta})$, if there is no quantum noise, the output quantum state is 
\begin{equation}
|\phi^{ideal}_{out}\rangle=U(\vec{\gamma},\vec{\beta})|\phi_{in}\rangle.
\end{equation}
We use $N$ to denote the number of quantum gates in the QAOA quantum circuit. The noisy output quantum state is written
\begin{equation}
\rho_{out}^{noise}=\sum_{i_1,i_2,...,i_N}{p_{1,i_1}p_{2,i_2}...p_{N,i_N}|\phi_{i_1,i_2,...,i_N}\rangle\langle\phi_{i_1,i_2,...,i_N}|},
\end{equation}
where $i_j=0,1,2,...,s$, $p_{k,i_j}$ signifies the probability that the $i_j$th noise Kraus operator acts when the quantum state evolves to the $k$-th quantum gate. The fidelity between $|\phi_{out}^{ideal}\rangle$ and $\rho_{out}^{noise}$ is 
\begin{equation}
F=\langle \phi_{out}^{ideal}|\rho_{out}^{noise}|\phi_{out}^{ideal}\rangle.
\end{equation}
The proportion of the ideal output in the noisy output is $p_{1,0}p_{2,0}...p_{N,0}=(1-p)^N$, where $1-p=a_0^2$. Except for $|\phi_{0,0,...,0}\rangle$, only a small part of $|\phi_{i_1,i_2,...,i_N}\rangle$ has a value for $|\langle \phi^{ideal}_{out}|\phi_{i_1,i_2,...,i_N}\rangle|^2$ that is significantly larger than 0. Considering this part of $|\phi_{i_1,i_2,...,i_N}\rangle$, we infer the output state fidelity is expressed as
\begin{equation}\label{EQ_state_fidelity}
F=(1-p)^{\delta N},
\end{equation}
for which $\delta$ is a constant that depends on the quantum circuit architecture and quantum noise model.

\subsection{Cost function}\label{Inf_on_cost}

The QAOA cost function is defined in Eq.(\ref{cost_function}). In Max-Cut problem, the problem Hamiltonian $H_p$ is written as
\begin{equation}
H_p=\sum_{i,j}{-C_{ij}\frac{1-Z_iZ_j}{2}}.
\end{equation}
In our work, we redefined $H_p$; specifically, we discarded its constant term and multiplied the result by 2 yielding
\begin{equation}
H_p=\sum_{i,j}{C_{ij}Z_iZ_j}.
\end{equation}
The performance and results of the QAOA are not influenced. One feature of the redefined $H_p$ is that its statistical average $\langle H_p \rangle$ in the entire space satisfies 
\begin{equation}
  \langle H_p \rangle=Tr(\frac{H_p}{2^m})=0,
\end{equation}
where $m$ denotes the number of qubit.

For a specific QAOA quantum circuit $U(\vec{\gamma},\vec{\beta})$ with $N$ quantum gates, the cost function is, if there is no quantum noise,
\begin{equation}
f(\vec{\gamma},\vec{\beta})^{ideal}=\langle\phi_{out}^{ideal}|H_p|\phi_{out}^{ideal}\rangle.
\end{equation}
However, as shown in Sec.~\ref{Inf_on_state}, when quantum noise $K$ acts on the QAOA quantum circuit, the output state density matrix changes. The noisy cost function is then written
\begin{equation}
\begin{split}
&f(\vec{\gamma},\vec{\beta})^{noise}=\\
&\sum_{i_1,i_2,...,i_N}{p_{1,i_1}p_{2,i_2}...p_{N,i_N}
\langle\phi_{i_1,i_2,...,i_N}|H_p|\phi_{i_1,i_2,...,i_N}\rangle}.
\end{split}
\end{equation}

We classify $|\phi_{i_1,i_2,...,i_N}\rangle$ in the following way: $|\psi_k\rangle$ represents $|\phi_{i_1,i_2,...,i_N}\rangle$, for which the number of $i_j\not=0$ is $k$, $k=0,1,2,...,N$. Each $|\psi_k\rangle$ contains $C_N^k$ members. When $k$ is small, $\langle\psi_k|H_p|\phi_k\rangle$ is close to the ideal cost function. As $k$ increases, $\langle\psi_k|H_p|\phi_k\rangle$ approaches the average of the arbitrary state. We use $(1-p)^{\alpha N} f(\vec{\gamma},\vec{\beta})^{ideal}$ to represent the value of $f(\vec{\gamma},\vec{\beta})^{noise}$ when $k$ is small; here, $\alpha$ is a constant number. Therefore, the effect of the QAOA on the cost function is expressed as
\begin{equation}
f(\vec{\gamma},\vec{\beta})^{noise}=(1-p)^{\alpha N}
f(\vec{\gamma},\vec{\beta})^{ideal}+(1-((1-p)^{\alpha N}))A
\end{equation}
in which $A$ represents the average cost function of an arbitrary state. As shown before, $A=Tr(\frac{H_p}{2^m})=0$.

From the above formula, we find that noise does not change the architecture of the parameter space but flattens the parameter space, flattening factor being $(1-p)^{\alpha N}$, where $\alpha$ depends on the quantum circuit architecture and the quantum noise model adopted.

\subsection{Cost function gradient}\label{NIGC}

Parameter optimization is an important procedure in QAOA. We use a classical algorithm to optimize the QAOA quantum circuit parameters. There are two such algorithms: one gradient-free and the other gradient-based. When we use the gradient-based optimization algorithm, we need to compute the cost function gradient. Quantum noise has an effect on the cost function and therefore affects the gradient of the cost function. In this section, we investigate this effect. 

We use the method given in \cite{chen_vqnet_2019} to compute the cost function gradient. A QAOA quantum circuit is governed by parameters, $\vec{\gamma}$ and $\vec{\beta}$ with respect to which the cost-function derivatives are written
\begin{equation}
\frac{\partial{ f(\vec{\gamma},\vec{\beta}) }}{\partial{\gamma_k}}
=\sum_{i,j} {-2C_{ij}\frac{f(\vec{\gamma},\vec{\beta})_{kij}^+
-f(\vec{\gamma},\vec{\beta})_{kij}^-}{2}},
\end{equation}
\begin{equation}
\frac{\partial{ f(\vec{\gamma},\vec{\beta}) }}{\partial{\beta_k}}=
\sum_{i=1}^{m} {-2\frac{f(\vec{\gamma},\vec{\beta})_{ki}^+-f(\vec{\gamma},\vec{\beta})_{ki}^-}{2}}
\end{equation}
In the first formula, $ f(\vec{\gamma},\vec{\beta})_{kij}^+ $ represents the cost function obtained by replacing $\gamma_k$, which is related to the $C_{ij}$ edge in the QAOA quantum circuit, with $\gamma_k+\pi/2$; $ f(\vec{\gamma},\vec{\beta})_{kij}^- $ is similar with the difference being that in the same position $\gamma_k$ is replaced by $\gamma_k-\pi/2$. Similarly, in the second formula, $ f(\vec{\gamma},\vec{\beta})_{ki}^+ $ represents the cost function obtained by replacing $\beta_k$, which is related to the $i$-qubit in the QAOA quantum circuit, with $\beta_k+\pi/2$; $ f(\vec{\gamma},\vec{\beta})_{ki}^- $ represents the cost function obtained by replacing $\beta_k$, which is related to $i$-qubit in QAOA quantum circuit, with $\beta_k-\pi/2$. In the summation, $m$ is the number of qubits in the QAOA quantum circuit.

From these two formulas, we find the derivatives of the cost function with respect to the QAOA parameters are composed of a series of different cost functions. We have already discussed the effect of quantum noise on the cost function; the effect of quantum noise on the cost function gradient can then be written as
\begin{equation}
\frac{\partial{ f(\vec{\gamma},\vec{\beta}) }}{\partial{\theta_k}}^{noise}
=(1-p)^{\alpha n}\frac{\partial{ f(\vec{\gamma},\vec{\beta}) }}{\partial{\theta_k}}^{ideal}
\end{equation}
where $\theta_k\in\{\vec{\gamma},\vec{\beta}\}$.

Clearly, concerning quantum noise, these derivatives are proportional to the ideal derivatives of the cost function, the scale factor being $(1-p)^{\alpha N}$. For different $\theta_k$, the constant number $\alpha$ is different, but as the quantum circuit is similar, different $\alpha$s will be close, so $(1-p)^{\alpha N}$ is close for different $\theta_k$. With increasing parameter $p$, the absolute value of the derivative of cost function decreases, and hence the absolute value of the cost function gradient decreases. Moreover, the direction of the cost function gradient changes little.

Therefore, we draw the conclusion that quantum noise only slows the parameter optimization speed. As the parameter optimization direction does not change, the optimized parameter values are similar in both noisy and noise-free conditions.

\section{Numerical Simulation}
\begin{figure}[htbp]
  \centering
  \includegraphics[width=8.6cm]{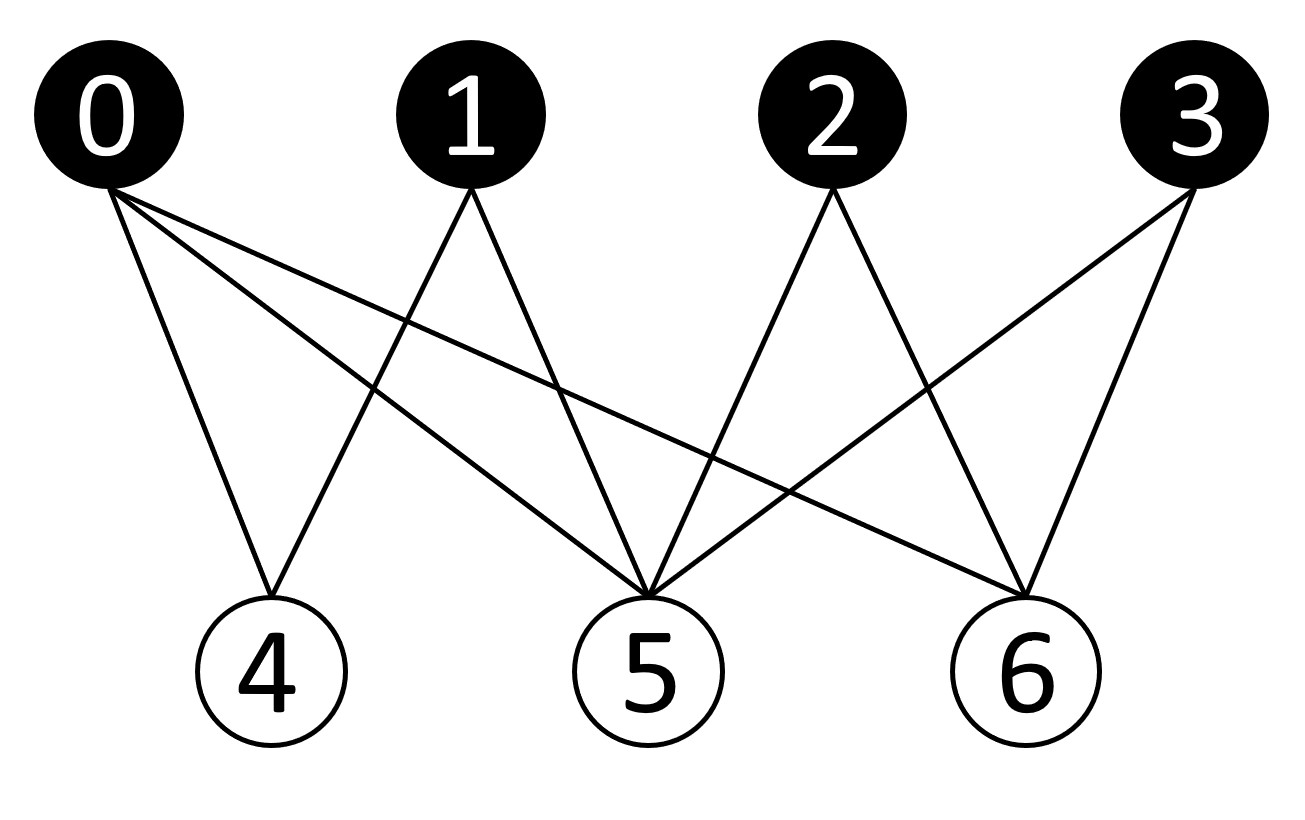}
  \caption{Max-Cut graph}
  \label{MaxCutGraph}
\end{figure}

We demonstrate our conclusions described in Sec.~\ref{EFFECT_ON_QAOA} with some numerical simulations. The specific Max-Cut problem we treat corresponds to finding the maximum cut in an undirected graph. The undirected graph information of the Max-Cut problem we selected is shown in Fig.~\ref{MaxCutGraph} and Table.~\ref{edgeweight}.

\begin{table}
  \tiny
	\caption{Edge weight table}
	\resizebox{8.6cm}{!}{ \begin{tabular}{|c|c|c|c|}
	\hline 
	Edge&Weight&Edge&Weight \\
	\hline  
	0,4&0.73&2,5&0.88\\
	\hline 
	0,5&0.33&2,6&0.58\\
	\hline 
	0,6&0.50&3,5&0.67\\
	\hline 
	1,4&0.69&3,6&0.43\\
  \hline
  1,5&0.36& & \\
	\hline  
	\end{tabular} }
	\label{edgeweight}
\end{table}

\subsection{State Fidelity}

\begin{figure*}[htbp]
  \centering
  \subfigure[Dephasing noise]{\includegraphics[width=5.2cm]{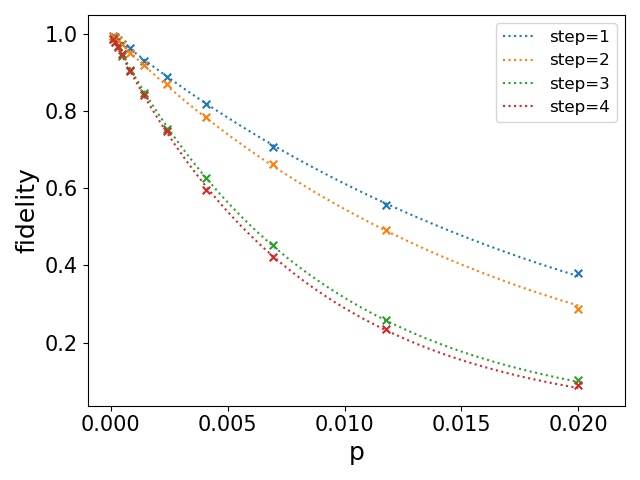}}
  \subfigure[Bit flip noise]{\includegraphics[width=5.2cm]{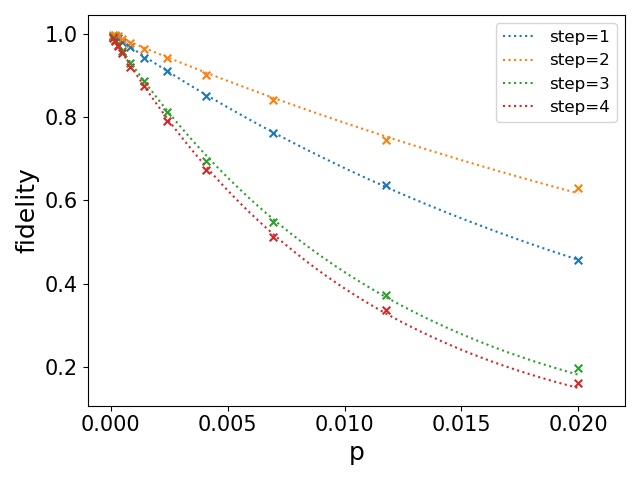}}
  \subfigure[Depolarizing noise]{\includegraphics[width=5.2cm]{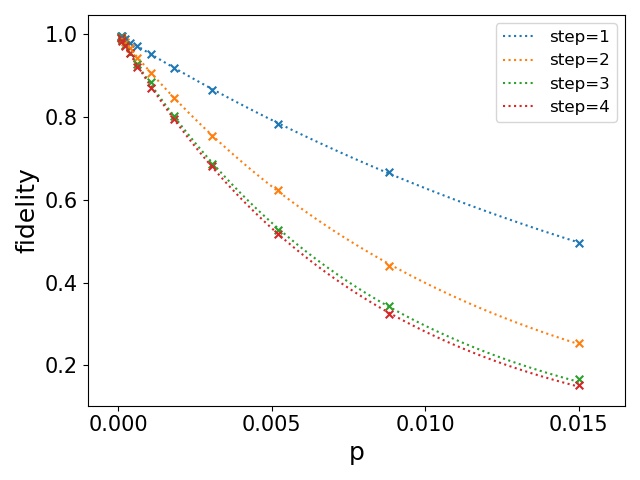}}
 
  \caption{Effect of quantum noise on output state fidelity. The horizontal axis is the noise parameter $p$, the vertical axis is the output state fidelity. QAOA step $n=1\sim4$. Panels (a),(b),(c) represent dephasing noise, bit flip noise, depolarizing noise respectively.}
  \label{Noisystatefidelity}
\end{figure*}

Firstly, we randomly set the QAOA quantum circuit parameters and compute the fidelity between ideal output state and noisy output state. The numerical results are shown in Fig.\ref{Noisystatefidelity}. It fits well with the Eq.(\ref{EQ_state_fidelity}).

\subsection{Cost Function}
\begin{figure*}[htbp]
  \centering
  \subfigure[Dephasing noise]{\includegraphics[width=5.2cm]
{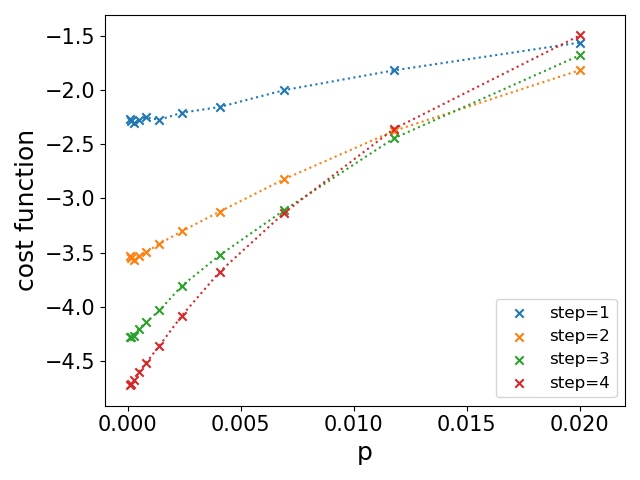}}
  \subfigure[Bit flip noise]{\includegraphics[width=5.2cm]
{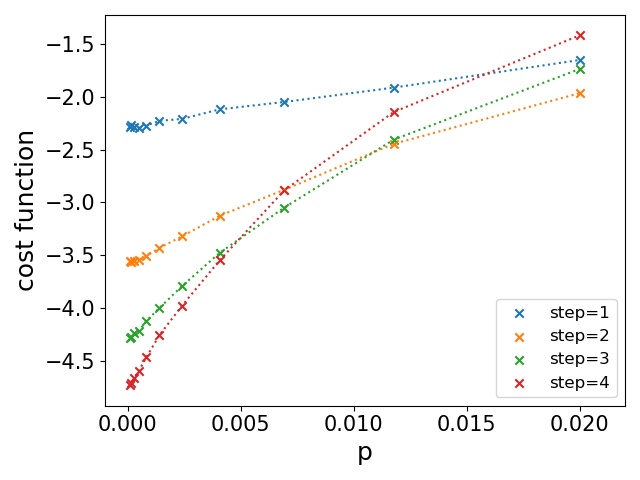}}
  \subfigure[Depolarizing noise]{\includegraphics[width=5.2cm]
{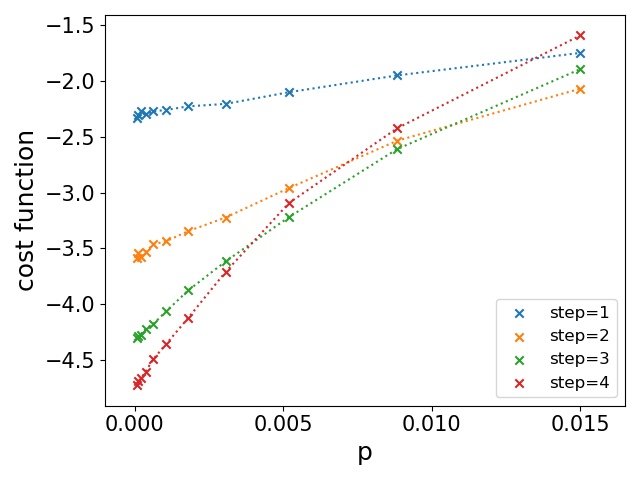}}
\caption{Effect of quantum noise on the cost function. The horizontal axis is the noise parameter $p$, the vertical axis is cost function $f(\vec{\gamma},\vec{\beta})$. Panels (a), (b), and (c) correspond to dephasing, bit-flip, and depolarizing quantum noise, respectively.}
  \label{Noisycostfunction}
\end{figure*}

Secondly, we demonstrate our conclusion proposed in Sec.~\ref{Inf_on_cost}. We set the QAOA quantum circuit parameters choosing values corresponding to the ideal optimal parameter settings. We compute the QAOA cost function by running the QAOA quantum circuit multiple times, in this process, statistical errors are inevitable. In Sec.~\ref{CFSE}, we discuss the statistical errors for the cost function in detail. We chose $M=5000$ in this test, from which the worst length of the $95\%$ confidence interval for the cost function is $L=0.051$. The simulation results are presented in Fig.~\ref{Noisycostfunction}. We found that when the quantum noise is small, the larger step $n$, the better the performance of the QAOA. On the other hand, the larger step $n$, the more the QAOA quantum circuit is affected by quantum noise. As the quantum noise parameter $p$ becomes larger, the performance improvement brought by QAOA step is offset by the effect of quantum noise. 

\begin{figure*}[htbp]
  \centering
  \subfigure[Dephasing noise]{\includegraphics[width=5.2cm]
{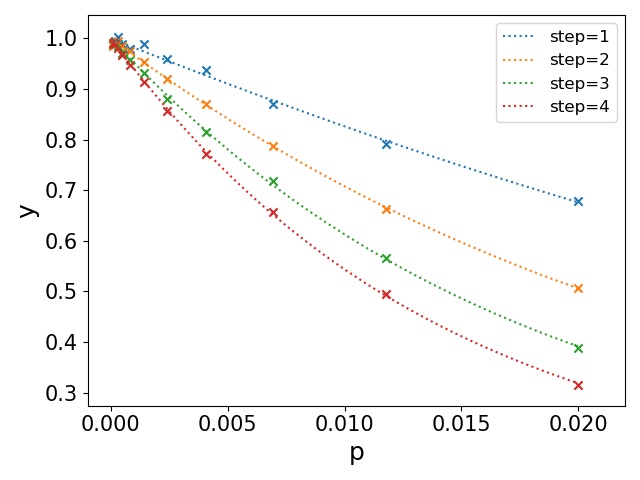}}
  \subfigure[Bit flip noise]{\includegraphics[width=5.2cm]
{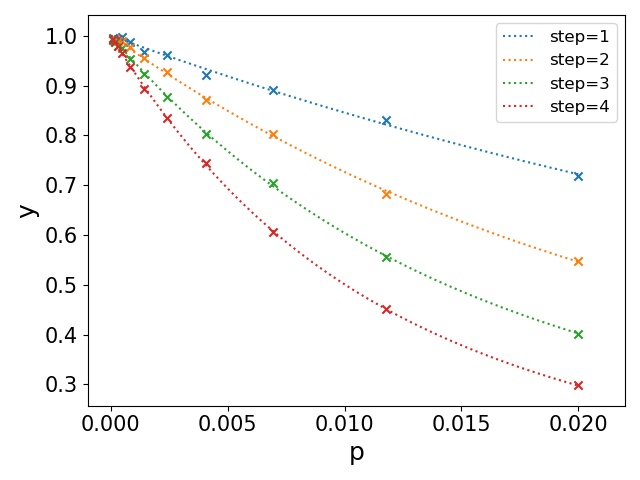}}
  \subfigure[Depolarizing noise]{\includegraphics[width=5.2cm]
{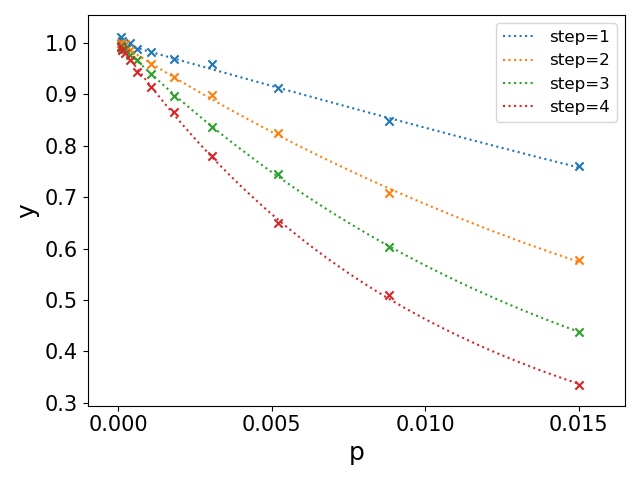}}
\caption{Data fitting showing the effect of quantum noise on the QAOA cost function in the test. The horizontal axis is the noise parameter $p$, the vertical axis is $y=\frac{f(\vec{\gamma},\vec{\beta})^{noise}}{f(\vec{\gamma},\vec{\beta})^{ideal}}$. Panels (a),(b), and (c) correspond to dephasing, bit flip, and depolarizing quantum noise, respectively.}
  \label{NoisycostfunctionPolyFitting}
\end{figure*}

We fit the data in the following way. We define
\begin{equation}
y=\frac{f(\vec{\gamma},\vec{\beta})^{noise}}{f(\vec{\gamma},\vec{\beta})^{ideal}},
\end{equation}
the relationship between $y$ and noise parameter $p$ being
\begin{equation}
y=(1-p)^{\alpha N}.
\end{equation}
We fit the experimental test data with this formula; the fitted curves are drawn in Fig.~\ref{NoisycostfunctionPolyFitting}.

\subsection{Cost Function Gradient}
Thirdly, We demonstrate our conclusion proposed in Sec.~\ref{NIGC}. In each simulation, we keep the QAOA quantum circuit parameter settings unchanged and assess the effect of quantum noise on the derivatives of the cost function with respect to parameters $\vec{\gamma}$ and $\vec{\beta}$. The statistical error of the cost function gradient caused by the measurement is discussed in Sec.~\ref{GSE}. Setting $M=5000$, the worst length of the $95\%$ confidence interval for $\frac{\partial f}{\partial \gamma}$ and $\frac{\partial f}{\partial \beta}$ are $L=0.130$ and $L=0.186$, respectively. When the absolute value of $\frac{\partial f}{\partial \gamma}$ or $\frac{\partial f}{\partial \beta}$ is small, the influence of the statistical error is considerable, so we choose the points for which the cost function gradient is large enough. The selected parameters are the parameters corresponding to the point where the cost function gradient is largest in the ideal optimization process.

\begin{figure*}[htbp]
  \centering
  \subfigure[Dephasing noise: $\frac{\partial f}{\partial \vec{\gamma}}$]
{\includegraphics[width=5.2cm]{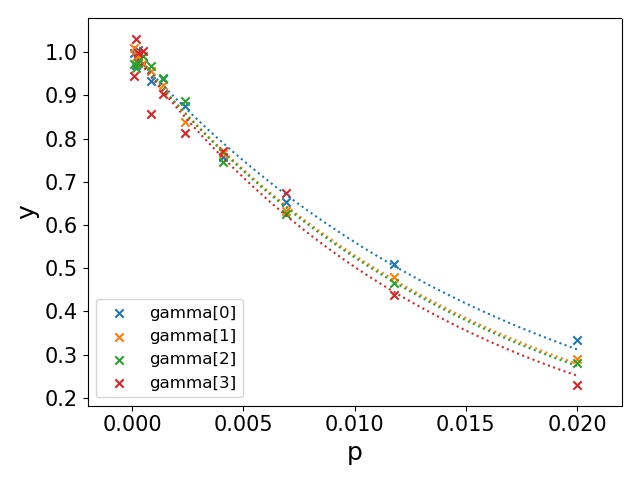}}
\subfigure[Bit flip noise: $\frac{\partial f}{\partial \vec{\gamma}}$]
{\includegraphics[width=5.2cm]{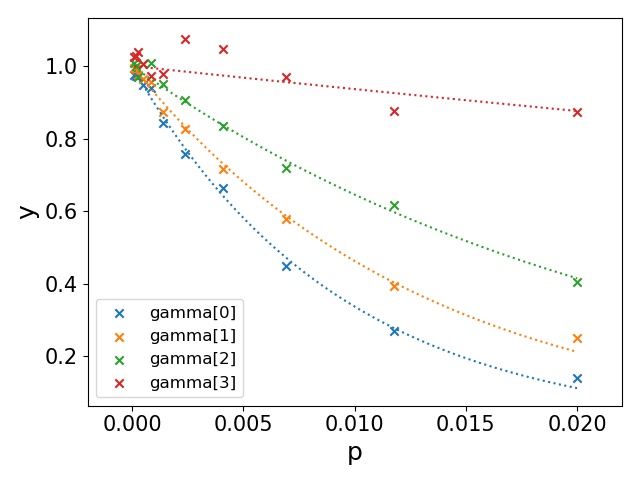}}
\subfigure[Depolarizing noise: $\frac{\partial f}{\partial \vec{\gamma}}$]
{\includegraphics[width=5.2cm]{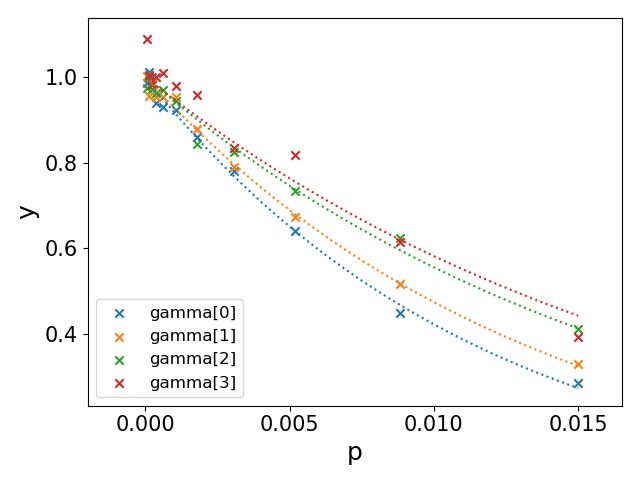}}
  \subfigure[Dephasing noise: $\frac{\partial f}{\partial \vec{\beta}}$]
{\includegraphics[width=5.2cm]{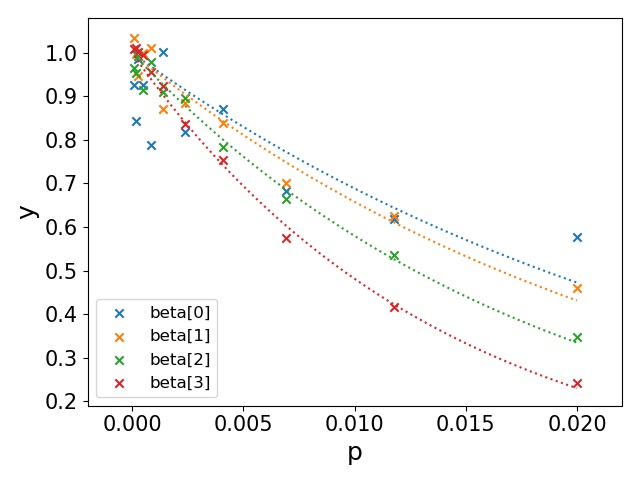}}
  \subfigure[Bit flip noise: $\frac{\partial f}{\partial \vec{\beta}}$]
{\includegraphics[width=5.2cm]{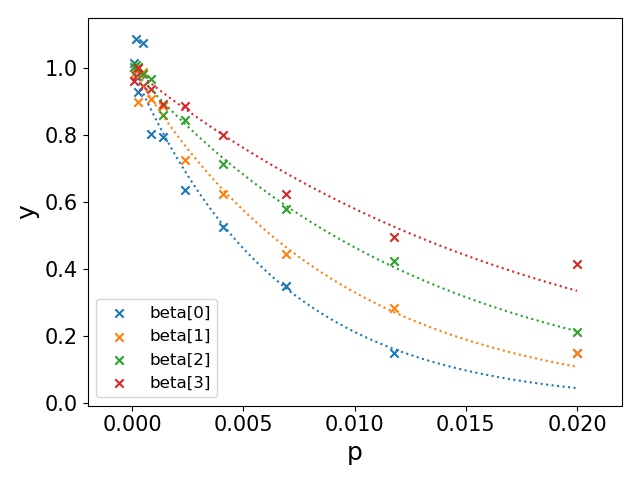}}
  \subfigure[Depolarizing noise: $\frac{\partial f}{\partial \vec{\beta}}$]
{\includegraphics[width=5.2cm]{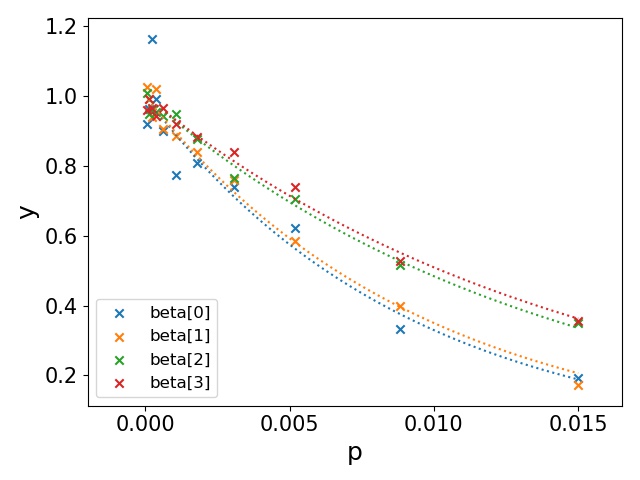}}
\caption{Relationship between the noise parameter $p$ and derivatives of the cost function with respect to the QAOA circuit parameters. Here, the QAOA step is $n=4$, and $y$ represents the ratio of the noisy derivative to ideal derivative. Panels (a) and (b) correspond to dephasing noise, (c) and (d) to bit-flip noise, and (e) and (f) to depolarizing noise.}
  \label{grad_test_fitting}
\end{figure*}

The result of the numerical simulation is displayed in Fig.~\ref{grad_test_fitting}. We only show the experimental data for step $n=4$, steps $n=1,2,3$ yielding similar data. For Fig.~\ref{grad_test_fitting}, we have defined 
\begin{equation}
y=\frac{\partial{ f(\vec{\gamma},\vec{\beta}) }}{\partial{\theta_k}}^{noise}/
\frac{\partial{ f(\vec{\gamma},\vec{\beta}) }}{\partial{\theta_k}}^{ideal}
\end{equation}
and fitted the experimental data with the approximate formula
\begin{equation}
y\approx(1-p)^{\alpha N}.
\end{equation}
There are four $\gamma$s and four $\beta$s, for which the values of the ideal derivatives of the cost function with respect to $\gamma$s are $[-4.30$,$-3.76$,$-2.87$,$-1.60]$, and the ideal derivatives of cost function with respect to $\beta$s are $[-1.31$,$-2.46$,$-3.40$,$-4.03]$. Consistent with our expectations, the fit for $\gamma[3]$ and $\beta[0]$ are poor, the cause being statistical errors. Nevertheless, the experimental data are consistent with our conclusions.


\subsection{Optimization Process}

Finally, we test the effect of quantum noise on the QAOA optimization process. We use the vanilla gradient descent algorithm to optimize the QAOA quantum circuit parameters. In the selected Max-Cut problem, we find that when the initial values of parameters $\vec{\gamma}$ and $\vec{\beta}$ are close to 0, we obtain the best results after parameter optimization. Hence, in our work, the initial values of $\vec{\gamma}$ and $\vec{\beta}$ are randomly chosen in the range $[-0.01,0.01]$.

\begin{figure*}[htbp]
  \centering
\subfigure[Dephasing]{\includegraphics[width=5.2cm]
{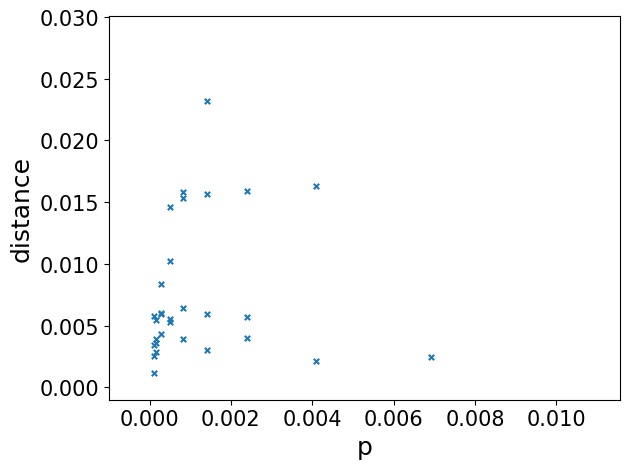}}
\subfigure[Bitflip]{\includegraphics[width=5.2cm]
{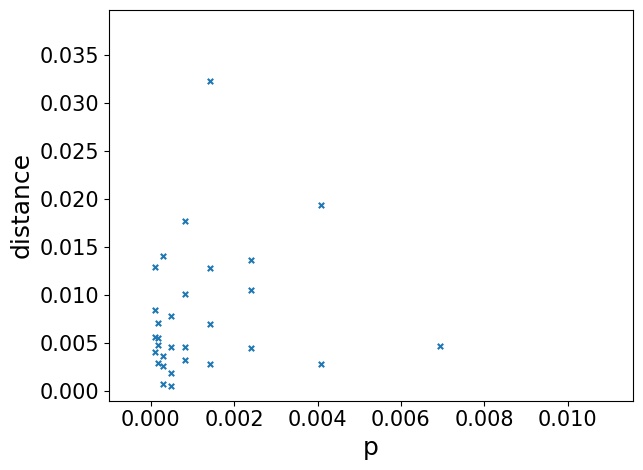}}
\subfigure[Depolarizing]{\includegraphics[width=5.2cm]
{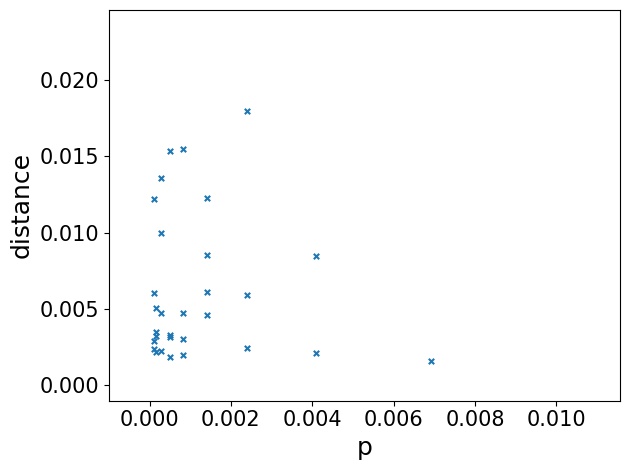}}
\caption{Distance between noisy optimized parameters and ideal optimized parameters. We chose points which satisfy $Np<0.5$. Panels (a), (b), and (c) correspond to dephasing, bit-flip, and depolarizing quantum noise, respectively.}
\label{parameter_distance}
\end{figure*}

We compute the distance between the ideal optimized parameters and the noisy optimized parameters. We use the root mean square of the Euclidean distance to describe the distance between two parameter sequences. Explicitly, we have
\begin{equation}
distance=\sqrt{\frac{|\vec{\gamma}^{noise}-\vec{\gamma}^{ideal}|^2
+|\vec{\beta}^{noise}-\vec{\beta}^{ideal}|^2}{2n}},
\end{equation}
where $2n$ represents the parameter number of QAOA quantum circuit. The result is shown in Fig.~\ref{parameter_distance}. The optimized parameters under quantum noise is close to the ideal optimized parameters. The difference arises from statistical errors. We infer that, if we set the execution times of QAOA quantum circuit large enough, the optimized parameters under quantum noise will be the same with the ideal optimized parameters. 

\section{Conclusion}

In summary, we investigated the effect of a class of quantum noise channels on QAOA performance and arrived at the following conclusions:
\begin{itemize}
\item [(1)] Quantum noise ﬂattens the QAOA parameter space, the factor being $(1-p)^{\alpha N}$.
\item [(2)] QAOA is a noise tolerant algorithm, quantum noise does not change the QAOA quantum circuit parameter optimization direction, for both noisy and ideal QAOA cost functions, the optimized QAOA parameters being nearly the same.
\end{itemize}

To support these conclusions, we evaluated the effects of dephasing, bit flip, and depolarizing quantum noise channels on the QAOA output, the test results being consistent with our conclusions. 

Our conclusions extend to other problems. The QAOA quantum circuit is a kind of multi-layer parameterized quantum circuit, and hence our conclusions apply equally to such circuits. 

There remain some unsolved problems. For example, our method is only suitable for certain quantum noise channels and is not applicable to arbitrary types. The effects of quantum noise on other NISQ quantum algorithms is also an open question. In the future, we will extend our method to account for these considerations and investigate the effects of quantum noise on other NISQ quantum algorithms.

\section*{Acknowledgments}
This work was supported by the National Key Research and Development Program of China (Grant No. 2016YFA0301700), the National Natural Science Foundation of China (Grants Nos. 11625419), the Strategic Priority Research Program of the Chinese Academy of Sciences (Grant No. XDB24030600)，and the Anhui Initiative in Quantum Information Technologies (Grants No. AHY080000).

\section*{Appendix}
\subsection{Simulation Method}

In our numerical simulations, we use a noisy quantum virtual machine (QVM) in the quantum programming architecture for the NISQ-device application (QPanda) as our noisy quantum simulator\cite{chen_qrunes_2019}. QPanda's noisy QVM supports all kinds of quantum noise models. Users can also define their own quantum noise model. QPanda also contains the variational quantum network (VQNet)---a framework to construct quantum-classical hybrid neural networks\cite{chen_vqnet_2019}. We use VQNet to realize the QAOA.

\subsubsection{Noisy Simulator}

First, we introduce the method of QPanda's noisy QVM. In noisy quantum circuits, the quantum state is a mixed state and represented by a density matrix. In QPanda's noisy QVM, we still use the state vector to represent the quantum state and combine the Monte Carlo method to simulate mixed-state evolution under noisy quantum operations\cite{noauthor_noise_nodate}. The specific method is as follows:
\begin{itemize}
\item [(1)] Assume the input state is a pure state denoted by $|\phi\rangle$ and $U$ a unitary quantum gate; the quantum noise is represented by Kraus operators $K=\{K_i\},i=1,2,...,s$.
\item [(2)] Compute $p_i=\langle\phi|K_i^{\dagger}K_i|\phi\rangle,i=1,2,...,s$. We note that $\sum_i{p_i}=1$.
\item [(3)] Generate a uniformly distributed random number $r$ in the range $[0,1)$, then find $l$ that satisfies 
\begin{equation}
\sum_{i=1}^{l-1}{p_i}\leq r\leq \sum_{i=1}^{l}{p_i}.
\end{equation}
Here we assume $ \sum_{i=1}^{0}{p_i}=0$.
\item [(4)] The expression for the evolution of state $|\phi\rangle$ is
\begin{equation}
|\phi_l \rangle \to U\frac{1}{\sqrt{p_l}}K_l|\phi\rangle.
\end{equation}
\item [(5)] Repeat procedure $(2)\sim(4)$ $M$ times. $M_l$ is the number of times that $K_l$ was selected. When $M$ is large enough, we know that $M_l\approx Mp_l$, and hence the output quantum state $\rho$ becomes
\begin{equation}
\rho\approx \sum_i{\frac{M_l}{M}|\phi_l\rangle \langle \phi_l|}.
\end{equation}
\end{itemize}

The noisy QVM realized by this method needs to run the quantum circuit multiple times. The density matrix of the output state may then be computed approximately from the statistical distribution of the results of multiple runs.

\subsubsection{VQNet}
\begin{figure}[htbp]
  \centering
    \includegraphics[width=8.6cm]{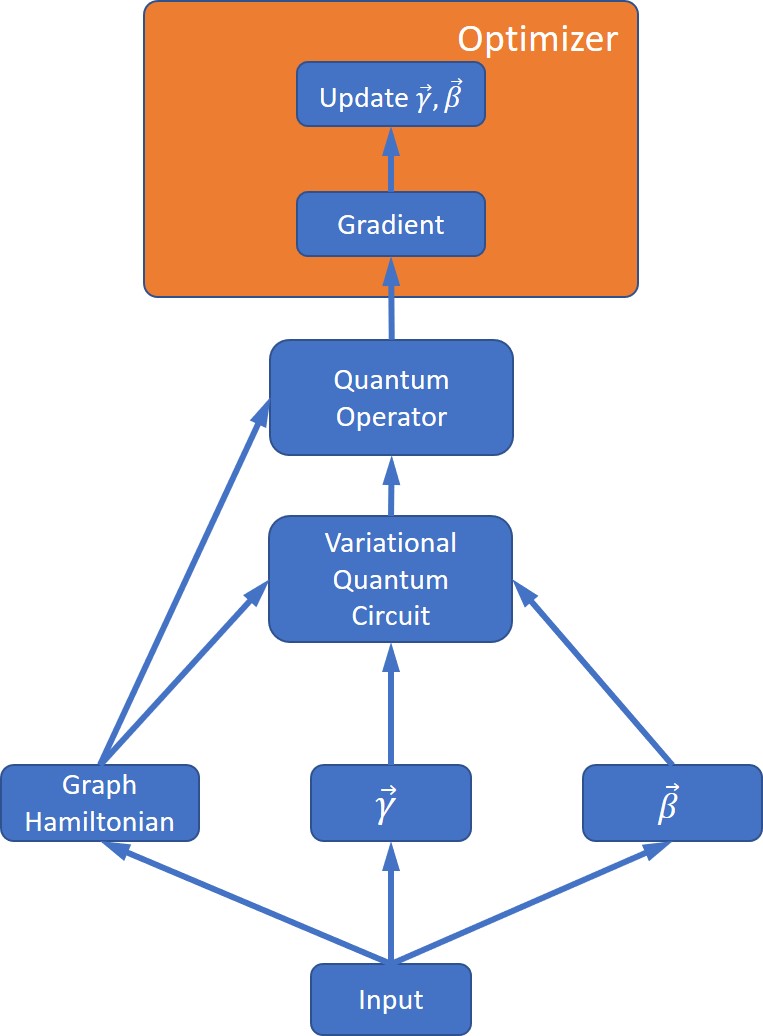}
    \caption{QAOA Flow Chart}
    \label{QaoaFlowChart}
\end{figure}

We use VQNet to realize QAOA. VQNet supports ideal and noisy QVM. The flow chart of QAOA is shown in Fig.~\ref{QaoaFlowChart}. $\vec{\gamma}$ and $\vec{\beta}$ are QAOA quantum circuit parameters, ``Variational Quantum Circuit'' represents the QAOA quantum circuit, ``Quantum Operator'' is a basic operator in VQNet, and the inputs of ``Quantum Operator'' are the ``Variational Quantum Circuit'' and the ``Graph Hamiltonian''. The output is the expectation of the input Hamiltonian after running the ``Variational Quantum Circuit''. The VQNet supports back propagation to get the derivatives of the cost function with respect to $\vec{\gamma}$ and $\vec{\beta}$. We then use the gradient descent algorithm to optimize parameters $\vec{\gamma}$ and $\vec{\beta}$.

\subsection{Statistical Error}\label{ASTError}

\subsubsection{Cost function statistical error}\label{CFSE}

To analyze cost function statistical error, we first introduce the method used to compute the cost function. The cost function is defined in Eq.(\ref{cost_function}). The method for calculating the expectation of $H_p$ is as follows: First, $H_p$ is decomposed into a sum of products of Pauli operators.
\begin{equation}
H_p=\sum_{i,j}{H_{ij}},
\end{equation}
where $H_{ij}=C_{ij}Z_iZ_j$. Then
\begin{equation}
\langle H_p\rangle=\sum_{i,j}{\langle H_{ij}\rangle}
\end{equation}
When we compute $\langle H_{ij}\rangle$, we run the QAOA quantum circuit multiple times, measuring qubit $i$ and qubit $j$. Second, we get the probabilities of the four quantum states $\{p^{ij}_{00},p^{ij}_{01},p^{ij}_{10},p^{ij}_{11}\}$, with $\langle H_{ij}\rangle$ becoming
\begin{equation}
\langle H_{ij}\rangle=C_{ij}(p^{ij}_{00}+p^{ij}_{11}-p^{ij}_{01}-p^{ij}_{10}).
\end{equation}
We define $p_{ij}=p^{ij}_{00}+p^{ij}_{11}$; the expectation of $H_p$ is then
\begin{equation}
\langle H_p\rangle=\sum_{i,j}{C_{ij}(2p_{ij}-1)}=B+\sum_{i,j}{2C_{ij}p_{ij}}
\end{equation}
where $B=-\sum_i{C_{ij}}$.

Third, we analyze the statistical error of the cost function. Consider the probability distribution, 
\begin{equation}
X=\{p_{ij}:1,1-p_{ij}:0\}.
\end{equation}
We define $X_{ij}=\sum_{i=1}^{i=M}{X_i}$, with $M$ representing the number of run times of the QAOA quantum circuit; the $X_i$s are mutually independent. When $M$ is large enough, $X_{ij}\sim N(M\mu_{ij},M\sigma_{ij}^2)$, where $N(M\mu_{ij},M\sigma_{ij}^2)$ represents a Gauss distribution with $\mu_{ij}=p_{ij}$ and $\sigma_{ij}=\sqrt{p_{ij}-p_{ij}^2}$. In terms of $X_{ij}$, the expectation value $\langle H_p\rangle$ is
\begin{equation}
\langle H_p\rangle=B+\sum_{i,j}{2C_{ij}\frac{X_{ij}}{M}}.
\end{equation}
Each $X_{ij}$ is also independent, therefore, $\langle H_p\rangle$ is a Gauss distribution taking the form
\begin{equation}
\langle H_p\rangle
=Y\sim N(B+\sum_{i,j}{2C_{ij}\mu_{ij}},4\sum_{i,j}{\frac{C_{ij}^2\sigma_{ij}^2}{M}}).
\end{equation}
Finally, we consider the $95\%$ confidence interval, which is written as $[\mu-\sigma,\mu+\sigma]$. The interval length is $4\sqrt{\sum_{i,j}{\frac{C_{ij}^2\sigma_{ij}^2}{M}}}$. When $p_{ij}=0.5$, $\sigma_{ij}$ has maximum value $\sigma_{ij}(max)=0.5$, hence the worst length of the $95\%$ confidence interval is 
\begin{equation}
L_{costfunction}=2\sqrt{\sum_{i,j}{\frac{C_{ij}^2}{M}}}.
\end{equation}
\subsubsection{Gradient Statistical Error}\label{GSE}

The formulas of the derivatives of the cost function to parameters $\vec{\gamma}$ and $\vec{\beta}$ are presented in \ref{NIGC}. They consist of some other cost functions, which were obtained by changing the QAOA quantum circuit parameter settings. Each cost function is a Gauss distribution and mutually independent. Therefore, the derivatives of the cost function with respect to parameters $\vec{\gamma}$ and $\vec{\beta}$ may also be regarded as Gauss distributions,
\begin{equation}
\frac{\partial{f(\vec{\gamma},\vec{\beta})}}{\partial{\gamma_k}}
\sim N(\mu_{\gamma_k},\sigma^2_{\gamma_k}),\\
\frac{\partial{f(\vec{\gamma},\vec{\beta})}}{\partial{\beta_k}}
\sim N(\mu_{\beta_k},\sigma^2_{\beta_k}),
\end{equation}
where
\begin{equation}
\mu_{\gamma_k}=\frac{\partial{f(\vec{\gamma},\vec{\beta})}}{\partial{\gamma_k}}^{ideal},
\sigma^2_{\gamma_k}=\sum_{i,j}{C_{ij}^2(\sigma_{kij+}^2+\sigma_{kij-}^2}),
\end{equation}
\begin{equation}
\mu_{\beta_k}=\frac{\partial{f(\vec{\gamma},\vec{\beta})}}{\partial{\beta_k}}^{ideal},
\sigma^2_{\beta_k}=\sum_{i=1}^{m}{(\sigma_{ki+}^2+\sigma_{ki-}^2}).
\end{equation}
As $\sigma^2_{\gamma_k}<2\sigma^2_{max}\sum_{i,j}{C_{ij}^2}$,$\sigma^2_{\beta_k} <2\sigma^2_{max}m$, therefore the worst $95\%$ confidence interval length of $\frac{\partial{f(\vec{\gamma},\vec{\beta})}}{\partial{\gamma_k}}$ and $\frac{\partial{f(\vec{\gamma},\vec{\beta})}}{\partial{\beta_k}}$ are
\begin{equation}
L_{\gamma_k}=2\sqrt{\frac{2}{M}}\sum_{i,j}{C_{ij}^2},\\
L_{\beta_k}=2\sqrt{2m\sum_{i,j}{\frac{C_{ij}^2}{M}}}.
\end{equation}

\bibliography{NoisyQAOA_pra}

\end{document}